# Effects of simultaneous real-time fMRI and EEG neurofeedback in major depressive disorder evaluated with brain electromagnetic tomography

Vadim Zotev[1#], Jerzy Bodurka[1,2#]

[1]Laureate Institute for Brain Research, Tulsa, OK, USA;
[2]Stephenson School of Biomedical Engineering, University of Oklahoma, Norman, OK, USA

**Abstract:** Recently, we reported an emotion self-regulation study (Zotev et al., 2020), in which patients with major depressive disorder (MDD) used simultaneous real-time fMRI and EEG neurofeedback (rtfMRI-EEG-nf) to upregulate two fMRI and two EEG activity measures, relevant to MDD. The target measures included fMRI activities of the left amygdala and left rostral anterior cingulate cortex, and frontal EEG asymmetries in the alpha band (FAA) and high-beta band (FBA). Here we apply the exact low resolution brain electromagnetic tomography (eLORETA) to investigate EEG source activities during the rtfMRI-EEG-nf procedure. The exploratory analyses reveal significant changes in hemispheric lateralities of upper alpha and high-beta current source densities in the prefrontal regions, consistent with upregulation of the FAA and FBA during the rtfMRI-EEG-nf task. Similar laterality changes are observed for current source densities in the amygdala. Prefrontal upper alpha current density changes show significant negative correlations with anhedonia severity. Changes in prefrontal high-beta current density are consistent with reduction in comorbid anxiety. Comparisons with results of previous LORETA studies suggest that the rtfMRI-EEG-nf training is beneficial to MDD patients, and may have the ability to correct functional deficiencies associated with anhedonia and comorbid anxiety in MDD.

## 1. Introduction

Simultaneous real-time fMRI and EEG neurofeedback (rtfMRI-EEG-nf) is an advanced neuromodulation technique that combines real-time fMRI neurofeedback (rtfMRI-nf) and EEG neurofeedback (EEG-nf) to enable simultaneous regulation of both hemodynamic (BOLD fMRI) and electrophysiological (EEG) brain activities (Zotev et al., 2014). The main promise of rtfMRI-EEG-nf for treatment of neuropsychiatric disorders is its ability to significantly alter both fMRI and EEG activity measures relevant to a specific disorder. However, implementation of rtfMRI-EEG-nf is technically challenging, and its mechanisms of action remain insufficiently investigated (see Lioi et al., 2020; Mano et al., 2017; Perronnet et al., 2017; Zotev et al., 2014, 2020).

Recently, we completed an emotion self-regulation study, in which patients with major depressive disorder (MDD) used rtfMRI-EEG-nf to simultaneously upregulate two EEG-nf and two rtfMRI-nf target measures, while inducing happy emotion (Zotev et al., 2020). The target measures for EEG-nf included right-vs-left frontal alpha EEG asymmetry (FAA) and left-vs-right frontal high-beta EEG asymmetry (FBA) for EEG channels F3 and F4. The rtfMRI-nf target measures included fMRI activity of the left amygdala (LA) and fMRI activity of the left rostral anterior cingulate cortex (L rACC). We selected these four target measures, because each of them is relevant to MDD (Zotev et al., 2020). During the rtfMRI-EEG-nf procedure, the MDD patients were able to significantly increase the FAA, the FBA, and the LA fMRI activation. They also achieved significant enhancement in fMRI functional connectivity between the LA and the L rACC through simultaneous upregulation of these regions' fMRI activities. After the rtfMRI-EEG-nf session, the MDD participants showed significant mood improvements, including reductions in state depression, anxiety, confusion, and total mood disturbance, and increase in state happiness (Zotev et al., 2020). These results demonstrated that the rtfMRI-EEG-nf may have potential for treatment of MDD.

The purpose of the present follow-up study is to evaluate EEG source activity during the rtfMRI-EEG-nf procedure. We employed the exact low resolution brain electromagnetic tomography (eLORETA) (Pascual-Marqui, 2007). The eLORETA method provides a distributed source solution to the electromagnetic inverse problem in neuroimaging, which enables exact, zero-error localization of point current sources in the presence of measurement and structured biological noise (Pascual-Marqui, 2007; Pascual-Marqui et al., 2011). The eLORETA solution is a linear, weighted minimum norm solution with low spatial resolution, meaning that reconstructed neighboring neuronal sources are highly correlated. The eLORETA analysis involves transformation of scalp EEG data to common average reference (Pascual-Marqui, 2007). In the LORETA-KEY software (The KEY Institute for Brain-Mind Research), the eLORETA solution space is restricted to cortical gray matter, including the hippocampus and the amygdaloid complex, and excluding other subcortical structures, such as the thalamus and the basal ganglia (see https://www.uzh.ch/keyinst/loreta.htm). The solution space is partitioned into 5×5×5 mm$^3$ voxels, and magnitude of current source density is computed for each voxel. It has been shown that the eLORETA method is superior to other EEG/MEG source analysis techniques in terms of localization accuracy and reliability of functional connectivity estimates (Pascual-Marqui et al., 2018). Results of LORETA source localization are generally consistent with activation patterns revealed by simultaneous fMRI (e.g. Mulert et al., 2004, 2005).

[#]Corresponding authors. E-mail: vzotev@laureateinstitute.org; jbodurka@laureateinstitute.org

We conducted the present study to investigate the following. *First*, we wished to examine to what extent the modulation of FAA and FBA by means of EEG-nf during the rtfMRI-EEG-nf procedure is reflected in hemispheric lateralities of cortical EEG sources. *Second*, we aimed to better understand potential therapeutic effects of the rtfMRI-EEG-nf by comparing our eLORETA results with findings from previous LORETA studies in MDD patients and healthy individuals. All analyses conducted in our work are exploratory, because the EEG-nf procedure targeted scalp EEG activity, while the eLORETA method estimates sources of neuronal activity.

## 2. Methods

### 2.1. Participants and procedures

The study, described in detail in Zotev et al., 2020, was conducted at the Laureate Institute for Brain Research. It was approved by the Western Institutional Review Board (IRB). Twenty four unmedicated MDD patients completed one rtfMRI-EEG-nf training session. The participants were right-handed and met the criteria for MDD laid out in the Diagnostic and Statistical Manual of Mental Disorders (DSM-IV, American Psychiatric Association, 2000). They underwent a psychological assessment, that included the Montgomery-Asberg Depression Rating Scale (MADRS, Montgomery and Asberg, 1979), the Snaith-Hamilton Pleasure Scale (SHAPS, Snaith et al., 1995), and other clinical ratings. The State-Trait Anxiety Inventory (STAI, Spielberger et al., 1970), the Profile of Mood States (POMS, McNair et al., 1971), and other tests were administered both before and after the session (Zotev et al., 2020).

During the session, participants in the experimental group (EG, *n*=16) were provided with the rtfMRI-EEG-nf, based on their real-time brain activity measures. Participants in the control group (CG, *n*=8) were provided, without their knowledge, with computer-generated sham feedback signals, unrelated to any brain activity (Zotev et al., 2018a, 2020).

The rtfMRI-EEG-nf was implemented using the custom real-time control system for integration of simultaneously acquired EEG and fMRI data streams (Zotev et al., 2014). The system utilizes real-time features of AFNI (Cox, 1996) and BrainVision RecView software (Brain Products, GmbH). The multimodal graphical user interface (mGUI) was used to display four variable-height bars (Fig. 1A). The bar heights, updated every 2 s, represented the four neurofeedback signals as follows. *First*, the magenta EEG-nf bar on the left represented a change in relative alpha EEG asymmetry for channels F3 and F4 (Fig. 2A, see e.g. Allen et al., 2001). The relative right-vs-left alpha asymmetry was defined as $A = (P(F4) - P(F3)) / (P(F4) + P(F3))$, where $P$ is EEG power in the alpha frequency band [7.5-12.5] Hz. In offline EEG analyses, normalized frontal alpha EEG asymmetry, commonly defined as FAA = $\ln(P(F4)) - \ln(P(F3))$, was employed. *Second*, the purple EEG-nf bar on the left represented a change in relative left-vs-right high-beta EEG asymmetry, defined as $B = (P(F3) - P(F4)) / (P(F3) + P(F4))$, where $P$ is EEG power in the high-beta (beta 3) frequency band [21-30] Hz. Normalized FBA = $\ln(P(F3)) - \ln(P(F4))$ was used in offline EEG analyses. *Third*, the red rtfMRI-nf bar on the right represented BOLD fMRI activity of the left amygdala (LA) target ROI (Fig. 2B). This spherical ROI with *R*=7 mm was centered at (−21, −5, −16) locus (Zotev et al., 2011) in the Talairach space (Talairach and Tournoux, 1988). *Fourth*, the orange rtfMRI-nf bar on the right represented fMRI activity of the left rACC target ROI (Fig. 2C). This ROI, also with *R*=7 mm, was centered at (−3, 34, 5) locus (Zotev et al., 2013). For a detailed explanation of selection of these four target measures, see Zotev et al., 2020.

The use of EEG-nf signals based on frontal hemispheric EEG power asymmetries, FAA and FBA, has important advantages over single-electrode (univariate) EEG-nf implementations. *First*, the asymmetry-based EEG-nf signals are less sensitive to variations in global EEG powers in the alpha and beta bands, reflecting changes in visual attention, arousal, mental concentration, etc. *Second*, such EEG-nf signals are less affected by EEG artifacts with extended topographies, such as MRI gradient-induced (MR) artifacts, cardioballistic (CB)

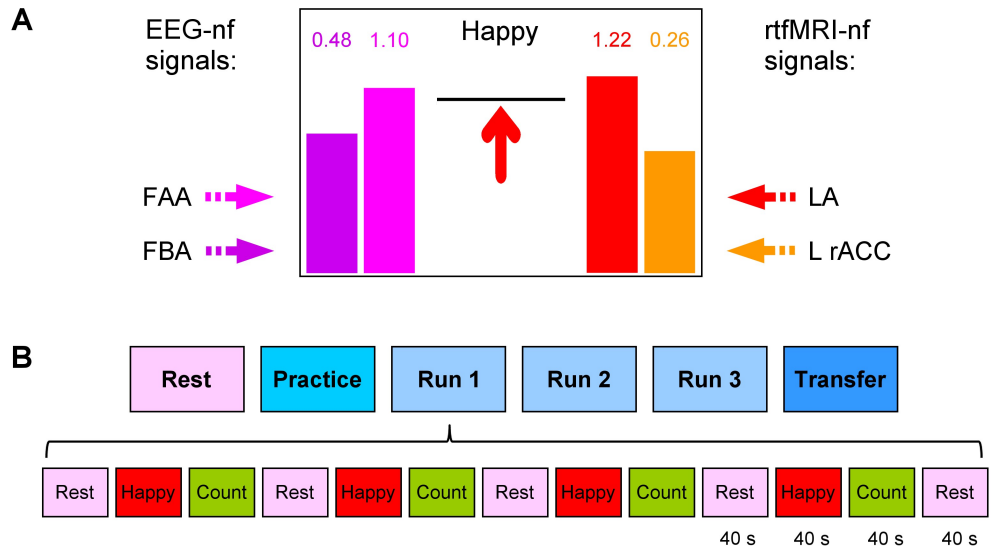

**Figure 1.** Experimental paradigm for emotion self-regulation training using simultaneous real-time fMRI and EEG neurofeedback (rtfMRI-EEG-nf). A) Real-time GUI display screen for Happy Memories conditions with rtfMRI-EEG-nf. The four neurofeedback signals are displayed on the screen as four variable-height bars. The two EEG-nf signals on the left are based, respectively, on changes in frontal alpha EEG asymmetry (FAA, magenta) and frontal high-beta EEG asymmetry (FBA, purple). The two rtfMRI-nf signals on the right are based, respectively, on fMRI activities of the left amygdala (LA, red) and the left rostral anterior cingulate cortex (L rACC, orange). The bar heights are updated every 2 s. B) Experimental protocol consists of six runs, each lasting 8 min 46 s. It includes a Rest run, four rtfMRI-EEG-nf runs – Practice, Run 1, Run 2, Run 3 – and a Transfer run without nf. The task runs consist of 40-s long blocks of Rest, Happy Memories, and Count conditions.

artifacts, and random-motion artifacts in simultaneous EEG-fMRI.

All study participants followed the experimental protocol depicted in Fig. 1B. It included six EEG-fMRI runs, each lasting 8 min 46 s. During the Rest run, the participants were asked to relax and rest while looking at a fixation cross. The five task runs – the Practice run, Run 1, Run 2, Run 3, and the Transfer run – consisted of alternating 40-s-long blocks of Rest, Happy Memories, and Count conditions (Fig. 1B). For the Rest conditions, the participants were instructed to relax and rest looking at a fixation cross. For the Happy Memories with rtfMRI-EEG-nf conditions, the participants were asked to induce happy emotion by retrieving happy autobiographical memories, and simultaneously raise the levels of all four neurofeedback bars (Fig. 1A). The bar heights represented changes in the target measures for the current Happy Memories condition (fMRI volume and EEG segment) relative to the baselines corresponding to the preceding Rest condition block. For the Count conditions, the participants were instructed to mentally count back from 300 by subtracting a given integer. The integers were 3, 4, 6, 7, and 9 for the five task runs, respectively. No bars were displayed during the Happy Memories conditions in the Transfer run, and during the Rest and Count conditions in all runs. For details of the experimental protocol and instructions given to the participants, see Zotev et al., 2020.

The experiments were performed using a General Electric Discovery MR750 3T MRI scanner with a standard 8-channel head coil. A single-shot gradient echo EPI sequence with 34 axial slices, *TR/TE*=2000/30 ms, and SENSE *R*=2 was employed for fMRI. It provided whole-brain fMRI images with $1.875 \times 1.875 \times 2.9$ mm$^3$ voxels. A T1-weighted 3D MPRAGE sequence yielded high-resolution anatomical brain images with $0.94 \times 0.94 \times 1.2$ mm$^3$ voxels.

EEG recordings were conducted simultaneously with fMRI using a 32-channel MR-compatible EEG system from Brain Products, GmbH. MR-compatible EEG caps (EASYCAP, GmbH) were custom modified to enable acquisition of four reference artifact waveforms for improved real-time EEG-fMRI artifact correction (Zotev et al., 2020). Because of this modification, the number of EEG channels was 27. Raw EEG data were acquired using BrainVision Recorder software (Brain Products, GmbH) with 0.2 ms temporal and 0.1 µV measurement resolution in [0.016-250] Hz frequency range with FCz reference. BrainVision RecView software was used to perform real-time EEG-fMRI artifact correction and export the corrected EEG data to the mGUI software for further processing. Technical details of the rtfMRI-EEG-nf implementation, real-time artifact correction, and real-time data processing were described previously (Zotev et al., 2014, 2020).

*2.2. EEG data processing*

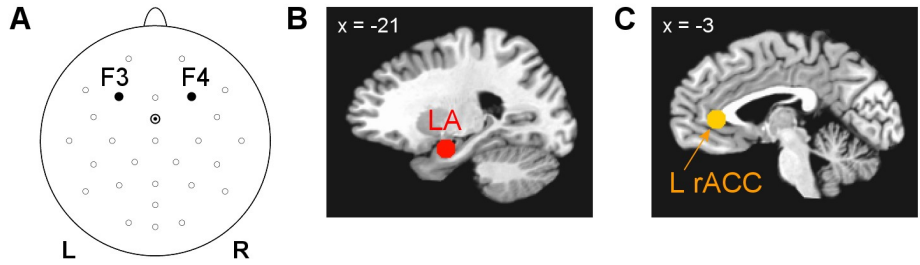

**Figure 2.** EEG channels and target regions of interest (ROIs) used to provide the rtfMRI-EEG-nf. A) Frontal EEG channels F3 (left) and F4 (right) used to generate EEG-nf signals based on frontal EEG asymmetries in the alpha and high-beta EEG bands. Channel FCz is the reference. B) Spherical target ROI for rtfMRI-nf in the left amygdala (LA) region. C) Spherical rtfMRI-nf target ROI in the left rostral anterior cingulate cortex (L rACC) region. The two ROIs, defined in the Talairach space, are transformed to each participants's individual fMRI image space.

Pre-processing of raw EEG data acquired during fMRI was performed using BrainVision Analyzer 2.1 software (Brain Products, GmbH), as described in Zotev et al., 2020. Briefly, it involved average artifact subtraction (AAS) for MR and CB artifacts and identification of bad intervals, showing intense random-motion artifacts. Independent component analysis (ICA) was applied to identify residual EEG-fMRI artifacts and various EEG artifacts, and to separate them from neuronal activity. The pre-processed EEG data included time courses of 27 EEG channels with single-electrode (FCz) reference and 250 S/s sampling (4 ms interval).

We defined the upper alpha EEG frequency band individually for each participant as [IAF, IAF+2] Hz, where IAF is the individual alpha peak frequency. The IAF was determined by inspection of average EEG spectra for the occipital and parietal EEG channels across the Rest condition blocks in the four nf runs (Fig. 1B). We focused on the upper alpha band, because EEG power changes in the lower (below IAF) and upper (above IAF) parts of the alpha band become increasingly dissociated as task demands increase: while the lower alpha activity reflects attentional demands, the upper alpha activity more specifically reflects cognitive and memory performance (e.g. Fink et al., 2005; Klimesch, 1999).

In preparation for the source analysis, the EEG data were transformed to common average reference, and the original reference was restored as the regular FCz channel, yielding 28 EEG channels. The data were lowpass filtered at 56 Hz (24 dB/octave), and downsampled to 125 S/s sampling rate (8 ms interval). The data for each task run were then split into three datasets, corresponding to the Happy Memories, Count, and Rest conditions (Fig. 1B), with exclusion of the bad intervals, and segmented into 4096-ms-long (512 time points) epochs.

## 2.3. eLORETA source analyses

The LORETA-KEY software was used to conduct the eLORETA source analyses. The software employs a realistic head model (Fuchs et al., 2002) and the probabilistic Montreal Neurological Institute (MNI) brain atlas (Mazziotta et al., 2001). The eLORETA transformation matrix was computed using MNI coordinates of the 28 EEG electrodes arranged according to the international 10-20 system (Fp1, Fp2, F3, F4, F7, F8, FC5, FC6, C3, C4, CP1, CP2, CP5, CP6, T7, T8, P3, P4, P7, P8, O1, O2, Fz, FCz, Cz, Pz, POz, Oz). To enable EEG source analysis in the frequency domain, we computed an EEG cross spectrum using each EEG channel's data for all the epochs corresponding to a given condition (Happy Memories, Count, Rest) in a given run. The cross spectra were calculated separately for the individual upper alpha band [IAF, IAF+2] Hz and for the high-beta band [21-30] Hz. The eLORETA transformation was then applied to the cross spectra, yielding current density magnitudes $j$ for $5\times5\times5$ mm$^3$ voxels ($n$=6239) as functions of frequency in the selected frequency bands. The eLORETA results were normalized as $\ln(j)$ without any additional scaling. Changes in the normalized current density values between the Happy Memories and Rest conditions were computed for each run for each participant.

Group statistical analyses were performed on changes in the normalized current density $\ln(j)$ between the two conditions. To evaluate task-specific activity, a single $t$-test relative to zero was applied to average individual current density changes in a given frequency band (upper alpha, high-beta). Correction for multiple comparisons across the eLORETA solution space was based on the randomization SnPM procedure (Nichols and Holmes, 2001), implemented in the LORETA-KEY software. The procedure yielded critical $t$-thresholds and corrected $p$-values. To evaluate correlations between localized activity and a psychological measure, a regression procedure was performed for average current density changes in a given frequency band and the psychological measure as an independent variable. The randomization SnPM procedure in this case yielded critical $r$-thresholds and corrected $p$-values.

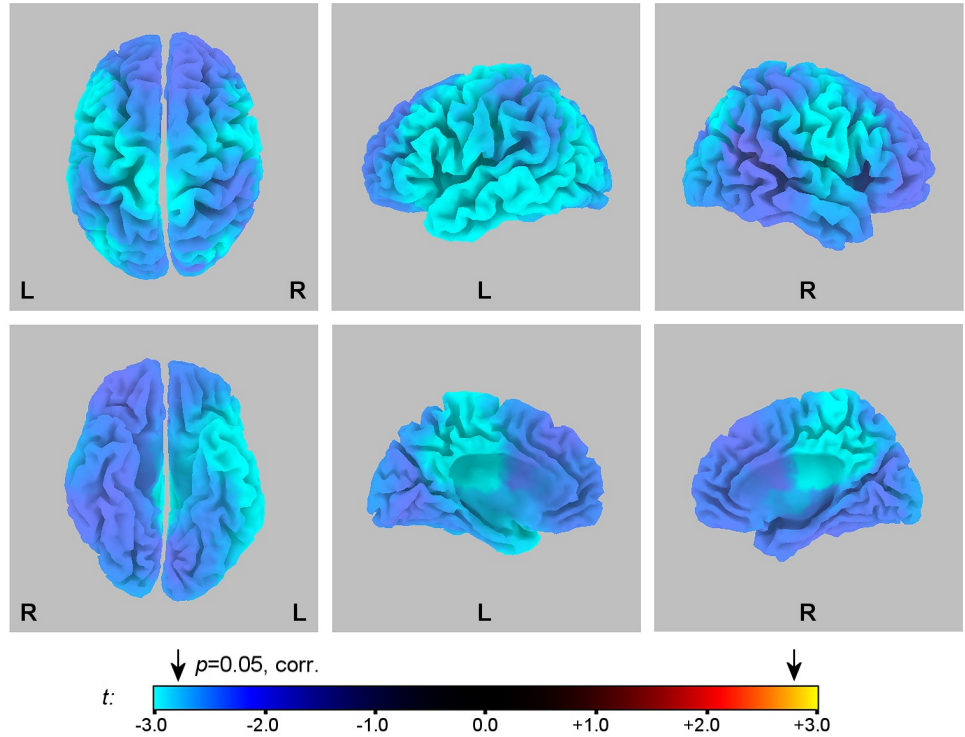

**Figure 3.** eLORETA statistical maps for changes in normalized upper alpha current source density ln(*j*) between the Happy Memories with rtfMRI-EEG-nf and Rest conditions (H vs R) for the experimental group (EG). Top row: views from the top, left, and right, respectively. Bottom row: a view from the bottom, a view of the left hemisphere from the medial plane, and a view of the right hemisphere from the medial plane, respectively. The blue/cyan color values denote reductions in the upper alpha current density during the Happy Memories with rtfMRI-EEG-nf conditions relative to the Rest conditions. The maps are projected onto the MNI152 template. The arrows above the colorbar designate the critical threshold from the randomization SnPM procedure. Statistics are summarized in Table 2.

## 2.4. Definitions of ROIs in the eLORETA space

To evaluate hemispheric laterality of eLORETA results, we defined, a priori, pairs of ROIs in the corresponding brain regions on the left and on the right. Because the EEG-nf signals were based on FAA and FBA for channels F3 and F4 (Fig. 2A), we chose ROIs in the middle frontal gyrus (MidFG) and superior frontal gyrus (SFG), located approximately underneath these two electrodes. Each ROI center was selected as a center of mass of an anatomical overlap between a gyrus and a Brodmann area (BA) in the same hemisphere, as defined in the Talairach-Tournoux Daemon (Lancaster et al., 2000). The centers of mass were determined using the 3dclust AFNI program with -mni option. The resulting ROI centers are specified in Table 1. The ROIs were then defined as collections of all eLORETA voxels within $R=10$ mm distance from the selected centers. In addition to the MidFG and SFG ROIs, we defined ROIs in the lateral orbitofrontal cortex, i.e. inferior frontal gyrus (IFG), BA 47, which plays an important role in emotion regulation, and in the amygdala (Table 1). The left amygdala ROI was centered at (−21, −4, −19) in MNI coordinates, corresponding to the center of the LA target ROI (Fig. 2B). The right amygdala ROI was centered at (21, −4, −19). Each ROI included approximately 15-20 voxels. Importantly, all the ROIs were defined independently of any results in the present study.

## 3. Results

### 3.1. eLORETA results for the upper alpha band

Figure 3 exhibits whole-brain statistical maps for changes in the normalized current density ln(*j*) in the upper alpha band during the Happy Memories with rtfMRI-EEG-nf conditions relative to the Rest conditions (H vs R) for the experimental group (EG). The upper alpha band was defined individually for each participant, as described above (Sec. 2.2). The individual-subject results were averaged across the four rtfMRI-EEG-nf runs (Practice, Run 1, Run 2, Run 3). The statistical results for voxels with $|t|>2.78$, two-tailed, are significant with $p<0.05$, corrected for multiple comparisons across the eLORETA space. The scale exponent for the color scale in Fig. 3 is 2. The most significant statistical results are reported in Table 2. For each anatomical area, the largest

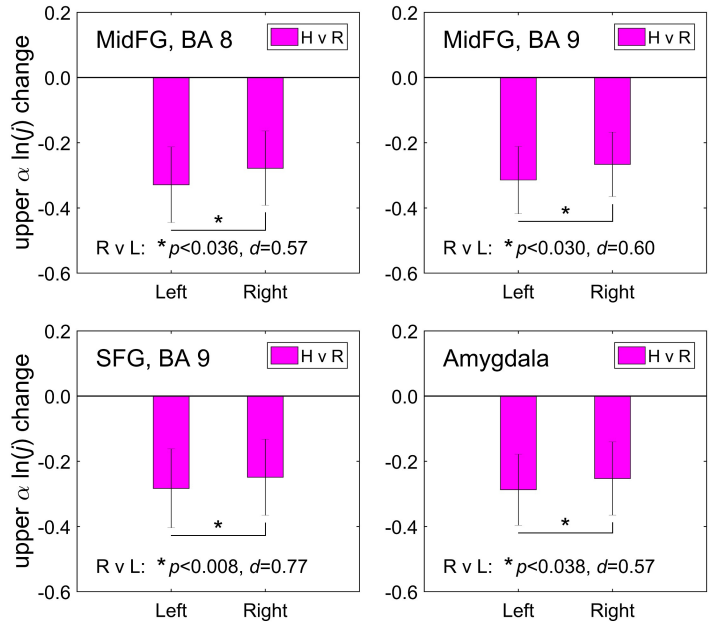

**Figure 4.** Laterality of the upper alpha current source density changes (Fig. 3) for corresponding ROIs on the left and on the right. The ROIs were selected a priori as described in the text. The statistics at the bottom of each figure (*p*-value from a paired *t*-test and effect size *d*) refer to comparison of the upper alpha current density changes on the right and on the left. MidFG – middle frontal gyrus, SFG – superior frontal gyrus.

Correlations between the changes in upper alpha current density and anhedonia severity

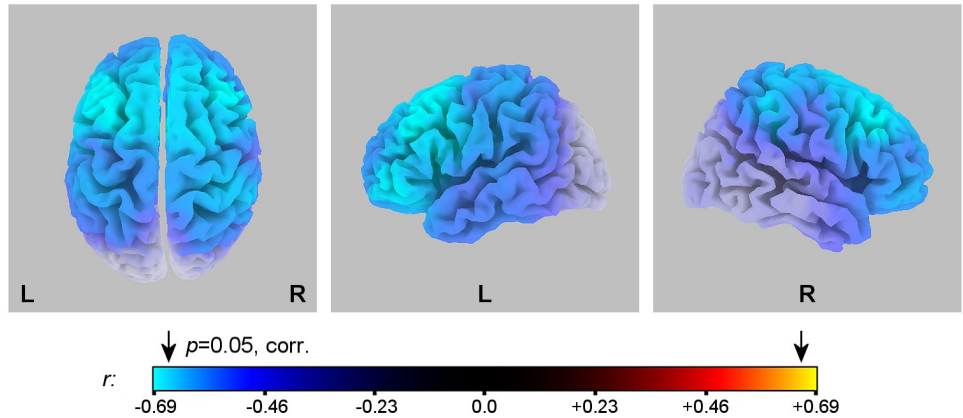

**Figure 5.** eLORETA statistical maps for correlations between the changes in normalized upper alpha current source density ln(*j*) for the Happy Memories with rtfMRI-EEG-nf conditions relative to the Rest conditions (H vs R) and anhedonia severity (SHAPS) ratings for the experimental group (EG). The blue/cyan color values mean greater reductions in the upper alpha current density, i.e. stronger activations, during the rtfMRI-EEG-nf conditions, relative to the Rest conditions, in patients with more severe anhedonia. The maps are projected onto the MNI152 template. The arrows above the colorbar denote the critical threshold from the randomization SnPM procedure. Statistics are summarized in Table 3. SHAPS – Snaith-Hamilton Pleasure Scale.

*t*-statistics value and the number of voxels in that area with statistics exceeding the critical threshold are specified in the table.

Figure 4 illustrates hemispheric laterality of the normalized upper alpha current density changes, exhibited in Fig. 3. Each sub-figure shows group mean (±sem) values for individual current density changes averaged within the corresponding ROIs on the left and on the right, defined a priori as described above (Sec. 2.4, Table 1).

Exploratory paired *t*-tests show that the reductions in upper alpha current density during the rtfMRI-EEG-nf task were significantly stronger, with medium effect sizes, for the prefrontal ROIs on the left, than for their counterparts on the right (MidFG, BA 8, R vs L: $t(15)=2.30$, $p<0.036$, $d=0.57$; MidFG, BA 9: $t(15)=2.40$, $p<0.030$, $d=0.60$; SFG, BA 9: $t(15)=3.08$, $p<0.008$, $d=0.77$). Significant positive change in the right-vs-left laterality of current densities is also observed for the amygdala ROIs (Amygdala, R vs L: $t(15)=2.27$, $p<0.038$, $d=0.57$). For the lateral orbitofrontal cortex ROIs, the laterality change trends toward significance (IFG, BA 47, R vs L: $t(15)=2.06$, $p<0.057$, $d=0.51$).

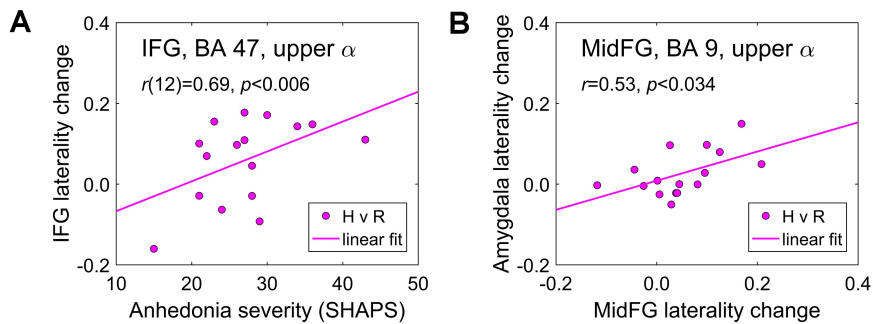

**Figure 6.** A) Correlation between the changes in right-vs-left upper alpha current density laterality for the a priori selected IFG (BA 47) ROIs during the rtfMRI-EEG-nf task and anhedonia severity (SHAPS) ratings for the experimental group (EG). B) Correlation between the changes in upper alpha current density lateralities for the amygdala ROIs and MidFG (BA 9) ROIs during the rtfMRI-EEG-nf task for the EG. IFG – inferior frontal gyrus.

The upper alpha current density changes are compared between the experimental and control groups (EG vs CG) in *Supplementary material* (S1.1, Fig. S1). The group differences did not reach statistical significance (supplementary Table S1). Changes in the right-vs-left lateralities of upper alpha current densities for the CG, corresponding to the laterality changes shown in Fig. 4 for the EG, were non-significant with small effects sizes.

Figure 5 shows statistical maps for correlations between the changes in the normalized upper alpha current density for the Happy Memories with rtfMRI-EEG-nf conditions relative to the Rest conditions (H vs R) and the EG participants' anhedonia severity (SHAPS) ratings. The voxels with $|r|>0.66$ correspond to $p<0.05$, corrected. Statistics for the results in Fig. 5 are included in Table 3. The negative correlations in Fig. 5 mean that the patients with more severe anhedonia (higher SHAPS) exhibited greater reductions in the upper alpha current density, i.e. stronger cortical activations, during the rtfMRI-EEG-nf task, relative to the Rest condition.

Correlations of the eLORETA results for the upper alpha band with several other psychological measures for the EG are reported in *Supplementary material* (S1.1, Figs. S2-S5). The correlations were not significant after the whole-brain correction. Note that there were strong associations among the depression severity (MADRS), anhedonia severity, and trait anxiety (STAI-t) ratings for the EG participants (MADRS vs SHAPS: $r=0.67$, $p<0.005$; SHAPS vs STAI-t: $r=0.75$, $p<8.7e-4$; MADRS vs STAI-t: $r=0.78$, $p<3.5e-4$). Similar to the results in Fig. 5, the upper alpha current density changes for the prefrontal and cingulate regions showed negative correlations with the patients' depression severity (MADRS, Fig. S2) and anxiety severity (STAI-t, Fig. S4).

Activations of these regions, i.e. reductions in their upper alpha current densities, were associated with mood improvements (S1.1, Figs. S3, S5).

Figure 6A illustrates an association between the changes in prefrontal upper alpha current density laterality during the Happy Memories with rtfMRI-EEG-nf conditions and the EG participants' anhedonia severity (SHAPS) ratings. The right-vs-left lateralities were computed for the a priori selected ROI pairs (Table 1). Partial correlation, controlled for the average current density changes for the left and right ROIs and for the patients' trait anxiety ratings (STAI-t), trended toward significance for the MidFG, BA 8 (Laterality change vs SHAPS: $r(12)=0.47$, $p<0.088$), and was significant for the IFG, BA 47 ($r(12)=0.69$, $p<0.006$) (Fig. 6A). For the amygdala ROIs, the partial correlation was also positive, but not significant ($r(12)=0.37$, $p<0.191$). The positive correlation means that MDD patients with more severe anhedonia (higher SHAPS) showed stronger reductions in upper alpha current density, i.e. stronger activations, for the ROI on the left, compared to its counterpart on the right.

Figure 6B demonstrates a connection between the changes in upper alpha current density laterality for the prefrontal regions and the corresponding laterality changes for the amygdala during the Happy Memories with rtfMRI-EEG-nf conditions for the EG. The right-vs-left lateralities were computed for the a priori selected ROIs (Table 1). The amygdala laterality changes exhibited significant across-subjects correlations with the MidFG, BA 9 laterality changes ($r=0.53$, $p<0.034$) (Fig. 6B), as well as with the IFG, BA 47 laterality changes ($r=0.79$, $p<3.0e-4$).

### 3.2. eLORETA results for the high-beta band

Figure 7 shows statistical maps for changes in the normalized current density $\ln(j)$ in the high-beta band between the Happy Memories with rtfMRI-EEG-nf conditions and Rest conditions (H vs R) for the experimental group (EG). The maps correspond to the first rtfMRI-EEG-nf run (Practice; average results across the four nf runs were not significant after the whole-brain correction). One case was identified as an outlier and excluded from the analysis (i.e. $n=15$ for the EG in this section), because of severe muscle artifacts, affecting reliability of estimated high-beta activity measures. The statistical results in Fig. 7 are significant ($p<0.05$) after the multiple comparisons correction for voxels with $|t|>3.03$, two-tailed. The scale exponent in Fig. 7 is 1. The statistical results are summarized in Table 4.

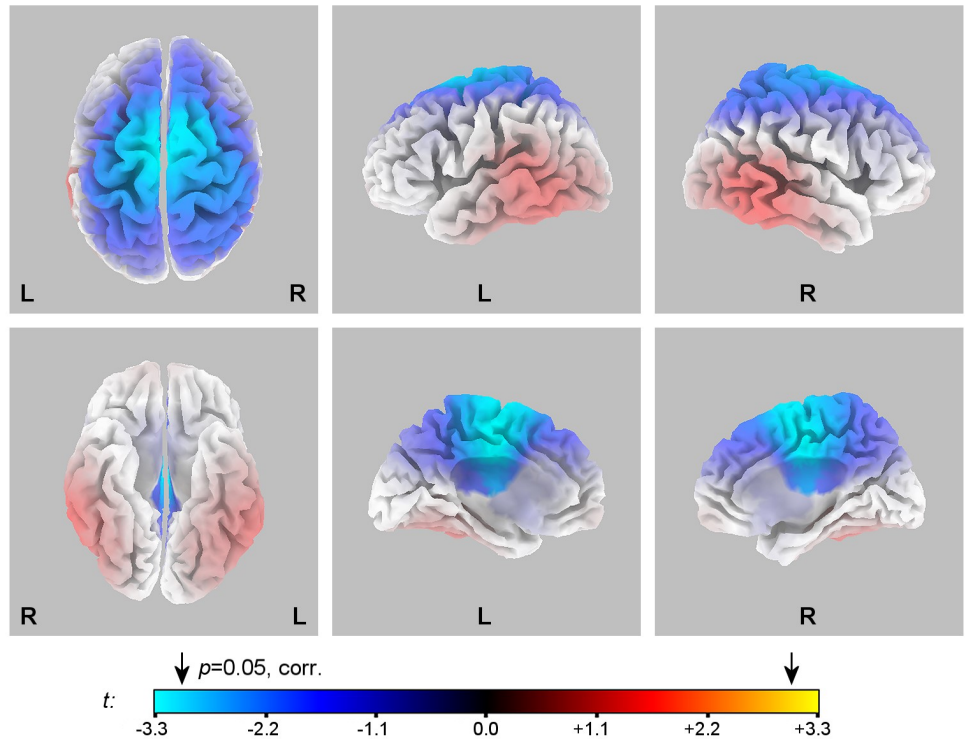

**Figure 7.** eLORETA statistical maps for changes in normalized high-beta (beta 3) current source density ln(*j*) between the Happy Memories with rtfMRI-EEG-nf and Rest conditions (H vs R) for the experimental group (EG). Top row: views from the top, left, and right, respectively. Bottom row: a view from the bottom, a view of the left hemisphere from the medial plane, and a view of the right hemisphere from the medial plane, respectively. The blue/cyan color values denote reductions in the high-beta current density during the Happy Memories with rtfMRI-EEG-nf conditions relative to the Rest conditions. The maps are projected onto the MNI152 template. The arrows above the colorbar designate the critical threshold from the randomization SnPM procedure. Statistics are summarized in Table 4.

Figure 8 reveals prefrontal hemispheric laterality of the normalized high-beta current density changes (Fig. 7). The individual current density changes were averaged within the a priori selected ROIs on the left and on the right (Table 1). The high-beta current density changes in Fig. 8 have smaller magnitudes than the corresponding changes for the upper alpha band in Fig. 4, because EEG power and associated current density are lower for the high-beta band, than for the alpha band. Exploratory paired *t*-tests indicate that the reductions in high-beta current density during the rtfMRI-EEG-nf task were significantly stronger, with medium or large effect sizes, for the SFG ROIs on the right, than for their counterparts on the left (SFG, BA 8, L vs R: $t(14)=2.22$, $p<0.043$, $d=0.57$; SFG, BA 9: $t(14)=3.35$, $p<0.005$, $d=0.86$). For the amygdala ROIs, the left-vs-right high-beta laterality change was also positive, but not significant.

The high-beta current density changes are compared between the experimental and control groups (EG vs CG) in *Supplementary material* (S1.2, Fig. S6). The group differences did not reach statistical significance (supplementary Table S6). Changes in the left-vs-right lateralities of high-beta current densities for the CG, corresponding to the laterality changes shown in Fig. 8 for the EG, were non-significant with small negative effects sizes.

Correlations of the high-beta eLORETA results with several psychological measures for the EG are reported in *Supplementary material* (S1.2, Figs. S7-S11). The correlations were not significant after the whole-brain correction. Notably, the high-beta current density changes for the right MidFG and SFG regions exhibited negative associations with the MDD patients' trait anxiety severity (STAI-t, Fig. S10). Right-lateralized reductions in high-beta activity along the cortical midline were associated with mood improvements (S1.2, Figs. S9, S11).

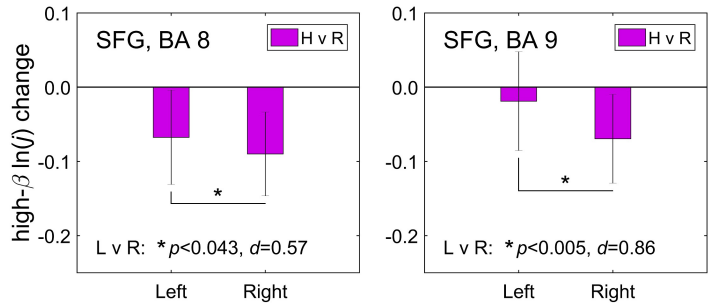

**Figure 8.** Laterality of the high-beta current source density changes (Fig. 7) for corresponding ROIs on the left and on the right. The ROIs were selected a priori as described in the text. The statistics at the bottom of each figure ($p$-value from a paired $t$-test and effect size $d$) refer to comparison of the high-beta current density changes on the left and on the right.

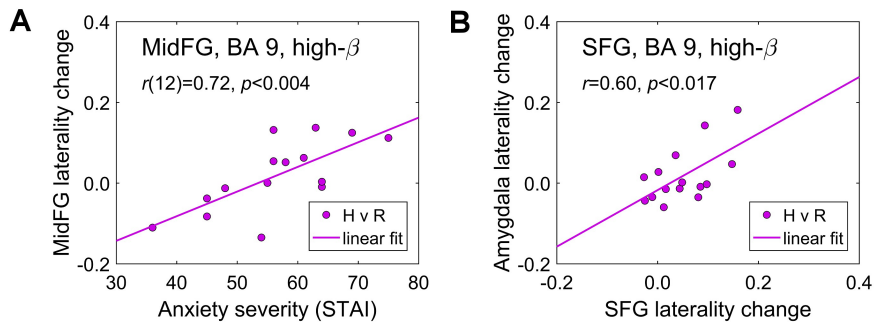

**Figure 9.** A) Correlation between the changes in left-vs-right high-beta current density laterality for the a priori selected MidFG (BA 9) ROIs during the rtfMRI-EEG-nf task and trait anxiety severity (STAI-t) ratings for the experimental group (EG). B) Correlation between the changes in high-beta current density lateralities for the amygdala ROIs and SFG (BA 9) ROIs during the rtfMRI-EEG-nf task for the EG. STAI – Stait-Trait Anxiety Inventory.

Figure 9A reveals an association between the prefrontal high-beta current density laterality changes during the Happy Memories with rtfMRI-EEG-nf conditions and the EG participants' anxiety severity (STAI-t) ratings. The left-vs-right lateralities were computed for the a priori selected ROI pairs (Table 1). Partial correlation, controlled for the average current density changes for the left and right ROIs, was significant for the MidFG, BA 8 (Laterality vs STAI-t: $r(12)=0.74$, $p<0.002$), and for the MidFG, BA 9 ($r(12)=0.72$, $p<0.004$) (Fig. 9A). For the amygdala ROIs, the partial correlation was also positive, but not significant ($r(12)=0.40$, $p<0.153$). The positive correlation means that MDD patients with more severe trait anxiety (higher STAI-t) showed stronger reductions in high-beta current density for the ROI on the right, compared to its counterpart on the left.

Figure 9B illustrates a connection between the changes in high-beta current density lateralities, during the Happy Memories with rtfMRI-EEG-nf conditions, for the prefrontal regions and for the amygdala. The left-vs-right lateralities were computed for the a priori selected ROIs (Table 1). Significant across-subjects correlations were found for the EG between the amygdala laterality changes and the corresponding laterality changes for several prefrontal ROI pairs, e.g. for the SFG, BA 9 ($r=0.60$, $p<0.017$) (Fig. 9B), the MidFG, BA 9 ($r=0.58$, $p<0.023$), and the IFG, BA 47 ($r=0.68$, $p<0.005$).

## 4. Discussion

In this paper, we reported the first application of EEG source analysis (eLORETA) to evaluate effects of simultaneous real-time fMRI and EEG neurofeedback. Our results show that the eLORETA analyses provide valuable new insights into mechanisms of rtfMRI-EEG-nf training, and complement the fMRI and EEG-fMRI analyses.

### 4.1. Discussion of the results for the upper alpha band

The eLORETA results for the upper alpha band (Fig. 3, Table 2) show widespread activation, indicated by reduction in the upper alpha current density, of the frontal, temporal, and parietal brain regions during the rtfMRI-EEG-nf task. These regions are consistent with those identified in the fMRI and EEG-fMRI analyses (Zotev et al., 2020), but with fewer activation centers detected in the limbic and sub-lobar regions (Table 2).

The results in Fig. 3 and Table 2 exhibit pronounced hemispheric laterality: reductions in the upper alpha current density during the rtfMRI-EEG-nf task are more significant for the frontal and temporal brain regions on the left than for the corresponding regions on the right. Within the prefrontal cortex, the most significant lateralized current density changes are observed for the left MidFG (BA 9, 8) and for the left IFG (BA 45, 47). The premotor cortex (precentral gyrus, BA 6) is activated bilaterally with similar lateralization. A LORETA study by Pizzagalli and colleagues demonstrated that reward responsiveness, a measure of approach motivation, is associated with reduced resting upper alpha (alpha 2) current density in the left prefrontal regions, including the left MidFG, SFG, and precentral gyrus (Pizzagalli et al., 2005). Similarly, an fMRI study by Spielberg and colleagues revealed an association between trait approach motivation and activation of the left MidFG during a cognitive task (Spielberg et al., 2011). Consistent with these findings, we interpret the left-lateralized activation of the MidFG and adjacent brain regions in our study as indicating enhanced approach motivation during the rtfMRI-EEG-nf task, as we argued previously (Zotev et al., 2016, 2020).

The eLORETA results also reveal significant activations in two adjacent areas of the left parahippocampal gyrus, corresponding to BA 34 and BA 28 (Table 2). Both activated areas include voxels in the left amygdala, targeted by the rtfMRI-nf in our study (Fig. 1A), more specifically − in its superficial (SF) subdivision. The SF amygdala subdivision is involved in reward processing and "modulation of approach-avoidance behavior in social interaction" (Bzdok et al., 2013). The observed SF activation is consistent with results of our previous studies that showed enhanced temporal correlations between the FAA in the upper alpha band and BOLD activity in the SF subdivision of the left amygdala during the rtfMRI-nf training (Zotev et al., 2016) and during the described rtfMRI-EEG-nf procedure (Zotev et al., 2020).

Importantly, the laterality effects in Fig. 3 are consistent with the significant positive changes in the FAA for channels F3 and F4 (Zotev et al., 2020), targeted with the EEG-nf in our study (Fig. 1A). This fact is illustrated in Fig. 4, which shows significant positive changes in right-vs-left lateralities of average current densities for the a priori selected MidFG and SFG ROIs, located approximately underneath EEG electrodes F3 and F4. A recent eLORETA study by Smith and colleagues demonstrated that the right-vs-left intracranial asymmetry scores for resting alpha current density are significantly lower in MDD patients than in healthy individuals for the MidFG and precentral gyrus (Smith et al., 2018). Similar lateralization was found for the upper alpha (alpha 2) band in an earlier LORETA study by Lubar and colleagues (Lubar et al., 2003). Our results in Fig. 4 indicate that such source asymmetry/laterality becomes more positive as MDD patients upregulate the FAA during the rtfMRI-EEG-nf procedure. Note that more approach-related emotional states are associated with more positive FAA levels (e.g. Stewart et al., 2014). Remarkably, the laterality of upper alpha current density for the amygdala ROIs is also significantly increased during the rtfMRI-EEG-nf task (Fig. 4).

The variations in upper alpha current density during the rtfMRI-EEG-nf procedure show significant negative associations with the participants' anhedonia severity ratings (Fig. 5, Table 3). These correlations are most pronounced for the prefrontal cortical areas, including the MidFG, the SFG, and the IFG (Table 3), and for the dorsal anterior cingulate cortex (dACC, BA 24, Table 3). The negative correlations mean that the MDD patients with higher anhedonia severity show stronger activations (greater reductions in upper alpha current density) of these brain areas during the rtfMRI-EEG-nf task. This result is consistent with findings reported in the literature. It has been observed that MDD patients, performing a cognitive task and achieving the same level of cognitive performance as healthy participants, exhibit higher activations of the dorsolateral prefrontal cortex (DLPFC) and dACC, than healthy participants (Fitzgerald et al., 2008; Harvey et al., 2005). The reason is that MDD patients need to recruit activities of these regions to a greater extent to achieve similar performance. Therefore, the results in Fig. 5, demonstrating positive associations between the anhedonia severity and the DLPFC and dACC activations, can be interpreted as showing the ability of the rtfMRI-EEG-nf to correct (i.e. reverse or normalize) functional deficiencies related to anhedonia. This interpretation was proposed and explained in our previous works (Zotev et al., 2016, 2018b, 2020). The correlation effects in Fig. 5 are most significant for the left MidFG (Table 3), suggesting that MDD patients with more severe anhedonia achieve greater enhancement in approach motivation during the rtfMRI-EEG-nf task. Therefore, the rtfMRI-EEG-nf procedure may be effective, in particular, at correcting approach motivation deficiencies associated with anhedonia in MDD.

The FAA changes for channels F3 and F4 during the rtfMRI-EEG-nf task exhibited significant positive association with the anhedonia severity ratings (Zotev et al., 2020). The changes in eLORETA upper alpha current density lateralities for the prefrontal ROI pairs also show positive correlations with the anhedonia severity ratings, though such correlations are less pronounced and become apparent when controlled for the trait anxiety severity (Sec. 3.1, Fig. 6A). This difference between the scalp-based and source-localized results can conceivably be attributed to effects of the common average reference transform, performed prior to the eLORETA analysis. Because the common average signal in the alpha band is dominated by posterior alpha rhythm, subtraction of this signal from the EEG data can obscure frontal alpha asymmetry effects (Hagemann et al., 2001). Large inter-subject variability in this signal's magnitude can increase variance in analyses of psychological associations of eLORETA activity measures. Nevertheless, the significant partial correlation between the current density laterality changes and anhedonia severity ratings (Sec. 3.1, Fig. 6A) suggests the potential of the rtfMRI-EEG-nf for correcting the prefrontal upper alpha source laterality deficits related to anhedonia.

The observation that the changes in upper alpha current density laterality for the prefrontal regions significantly correlate, during the rtfMRI-EEG-nf task, with the corresponding laterality changes for the amygdala (Sec. 3.1, Fig. 6B) is an important finding of the present study. It is in agreement with the significant association between the FAA changes and the amygdala BOLD activation laterality during the rtfMRI-nf training, we reported previously (Zotev et al., 2016). This finding shows that the upper alpha source laterality changes may be mutually consistent, under certain conditions, for the prefrontal regions and for the amygdala.

### 4.2. Discussion of the results for the high-beta band

The eLORETA results for the high-beta band (Fig. 7, Table 4) show pronounced reduction in high-beta activity of the fronto-centro-parietal cortical area, extending down to the cingulate gyrus, during the rtfMRI-EEG-nf task. We interpret this effect as an indication of reduction in anxiety. Indeed, the most common anxiety-related pattern revealed by LORETA is the elevated beta activity localized along the anterior cingulate or the midline cortex (Price and Budzynski, 2009). This includes elevated activity of this area in the high-beta band (e.g. Sherlin and Congedo, 2005; Velikova et al., 2010). A LORETA study by Pizzagalli and colleagues showed that MDD patients, compared to healthy individuals, exhibit increased resting high-beta current density in the right prefrontal regions, particularly the right SFG and IFG (Pizzagalli et al., 2002). The resting high-beta activity in those regions correlated with trait anxiety, accompanying depression (Pizzagalli et al., 2002). An sLORETA study by Paquette and colleagues demonstrated that alleviation of depressive symptoms after treatment is associated with reduction in high-beta current density in several brain regions on the right, including the right medial prefrontal cortex/dACC (BA 9/32) (Paquette et al., 2009). In the present study, the reductions in high-beta current density are more widespread for the fronto-centro-parietal regions on the right, than for those on the left (Fig. 7). This observation suggests that the changes in high-beta activity, achieved during the rtfMRI-EEG-nf procedure, may be beneficial to MDD patients.

Importantly, the laterality of the high-beta activity changes in Fig. 7 is consistent with the significant positive changes in the FBA for channels F3 and F4 (Zotev et al., 2020), upregulated using the EEG-nf (Fig. 1A). This is demonstrated in Fig. 8, which shows significant positive changes in the left-vs-right lateralities of average current densities for the a priori selected SFG ROIs. The positive laterality changes in this case are associated with greater reductions in the high-beta current densities for the right SFG regions, compared to the left SFG regions. The occurrence of these effects in the SFG, i.e. along the cortical midline, is consistent with the previous LORETA findings (Paquette et al., 2009; Pizzagalli et al., 2002). The positive high-beta laterality change for the amygdala ROIs is similarly associated with a stronger reduction in high-beta activity of the right amygdala.

The high-beta current density variations for the right MidFG and SFG regions during the rtfMRI-EEG-nf task show negative correlations with the MDD patients' anxiety severity ratings (supplementary Fig. S10). Though not significant, these negative correlations point to stronger reductions in high-beta activity for the right MidFG and SFG areas in MDD patients with higher trait anxiety. Therefore, the rtfMRI-EEG-nf may have the ability to correct (i.e. reduce or normalize) the abnormally elevated right-lateralized high-beta activity, which can be attributed to comorbid anxiety and avoidance motivation in MDD (e.g. Bruder et al., 2017; Trew, 2011).

Importantly, the changes in the left-vs-right laterality of high-beta current densities for the MidFG ROIs during the rtfMRI-EEG-nf task exhibit significant partial correlations with the anxiety severity ratings (Sec. 3.2, Fig. 9A). These results suggest the potential of the rtfMRI-EEG-nf for correcting the prefrontal high-beta source laterality deficits related to trait anxiety.

The changes in the high-beta current density laterality for the prefrontal regions show significant associations with the corresponding laterality changes for the amygdala during the rtfMRI-EEG-nf procedure (Sec. 3.2, Fig. 9B). This finding indicates that mutually consistent source laterality changes for the prefrontal regions and for the amygdala may be observed, under certain conditions, not only for the upper alpha band, but also for the high-beta band.

### 5. Study limitations

The main limitation of the present study, from the eLORETA source analysis perspective, is the relatively small sample size. This limitation resulted from the slow and difficult recruitment process for unmedicated MDD patients. While the eLORETA results for the experimental group (EG) were significant (Fig. 3 and Fig. 7), the differences between the experimental and control groups (EG vs CG) did not reach statistical significance (supplementary Fig. S1 and Fig. S6). Therefore, the reported eLORETA activity measures cannot be attributed specifically to the effects of the rtfMRI-EEG-nf for the EG, as opposed to those of the sham feedback for the CG. Nevertheless, the statistical maps for the EG vs CG group differences exhibit spatial patterns that are similar to those for the EG, both for the upper alpha band (Fig. S1 vs Fig. 3), and for the high-beta band (Fig. S6 vs Fig. 7). These spatial similarities suggest that the EG vs CG group differences primarily reflect the current density changes achieved by the EG participants using the rtfMRI-EEG-nf. The exploratory analyses conducted and the pilot results obtained can be very valuable and informative for designing future studies to replicate the reported findings and confirm their specificity to the rtfMRI-EEG-nf. Larger cohorts of participants could conceivably yield more significant experimental-vs-control group differences for eLORETA results. They would also allow more accurate evaluation of associations between eLORETA activity measures and various psychological metrics.

## 6. Conclusions

The eLORETA source analysis results, reported in this paper, lead to the following conclusions. *First*, performance of the rtfMRI-EEG-nf task is associated with significant positive changes in hemispheric lateralities of current source densities in the prefrontal cortical regions. These laterality changes are consistent with the significant positive changes in the upper alpha and high-beta frontal EEG asymmetries during the rtfMRI-EEG-nf task. *Second*, the EEG source activities during the rtfMRI-EEG-nf procedure are beneficial to MDD patients. Specifically, MDD patients with higher anhedonia severity demonstrate larger reductions in upper alpha current density in the left prefrontal regions, indicating an enhancement in approach motivation. MDD patients with higher anxiety severity exhibit larger reductions in high-beta current density in the right prefrontal regions, indicating a reduction in comorbid anxiety. These eLORETA findings suggest that the rtfMRI-EEG-nf may become an effective tool for treatment of major depression.

## 7. Availability of data

The data that support the findings and the data analysis scripts used in this study are available from the corresponding authors upon reasonable request.

**CRediT authorship contribution statement**

**VZ:** Conceptualization, Methodology, Investigation, Formal analysis, Writing – original draft, Writing – review & editing. **JB:** Conceptualization, Methodology, Project administration, Resources, Writing – review & editing.

**Declaration of competing interest**

The authors declare that they have no known competing financial interests or personal relationships that could have appeared to influence the work reported in this paper.

**Funding**

This work was supported by the Laureate Institute for Brain Research and the William K. Warren Foundation and in part by the P20 GM121312 award from the National Institute of General Medical Sciences, National Institutes of Health.

Supplementary data to this article can be found online at https://doi.org/10.1016/j.nicl.2020.102459

**Table 1.** Locations in the corresponding brain regions on the left and on the right used to define regions of interest (ROIs) in the eLORETA space. The ROIs were defined as collections of voxels within 10 mm distance from the specified centers.

| Region | Brodmann area | Left center $x, y, z$ (mm) | Right center $x, y, z$ (mm) |
| --- | --- | --- | --- |
| Middle frontal gyrus | 8 | −36, 24, 47 | 37, 24, 47 |
| Middle frontal gyrus | 9 | −43, 25, 37 | 45, 25, 37 |
| Superior frontal gyrus | 8 | −18, 32, 53 | 19, 32, 53 |
| Superior frontal gyrus | 9 | −20, 49, 36 | 22, 49, 36 |
| Inferior frontal gyrus | 47 | −39, 24, −11 | 39, 24, −11 |
| Amygdala | 34 | −21, −4, −19 | 21, −4, −19 |

$x, y, z$ – MNI coordinates.

**Table 2.** Changes in normalized upper alpha current source density between the Happy Memories with rtfMRI-EEG-nf and Rest conditions (H vs R) for the experimental group (EG).

| Region | Laterality | $x, y, z$ (mm) | $t$-score | Voxels |
| --- | --- | --- | --- | --- |
| **Frontal lobe** | | | | |
| Medial frontal gyrus (BA 6) | L,R | −10, −25, 50 | −3.41 | 68 |
| Precentral gyrus (BA 6) | L | −55, 0, 10 | −3.37 | 85 |
| Middle frontal gyrus (BA 9) | L | −45, 25, 40 | −3.30 | 31 |
| Inferior frontal gyrus (BA 45) | L | −55, 15, 5 | −3.30 | 28 |
| Inferior frontal gyrus (BA 47) | L | −50, 15, 0 | −3.29 | 77 |
| Precentral gyrus (BA 6) | R | 50, −5, 25 | −3.28 | 84 |
| Middle frontal gyrus (BA 6) | L | −30, −5, 55 | −3.27 | 63 |
| Precentral gyrus (BA 4) | L | −15, −30, 60 | −3.22 | 60 |
| Precentral gyrus (BA 4) | R | 50, −10, 45 | −3.19 | 38 |
| Middle frontal gyrus (BA 8) | L | −40, 25, 45 | −3.18 | 16 |
| Middle frontal gyrus (BA 6) | R | 40, −5, 50 | −3.15 | 53 |
| Middle frontal gyrus (BA 9) | R | 50, 25, 40 | −3.03 | 23 |
| **Temporal lobe** | | | | |
| Superior temporal gyrus (BA 22) | L | −50, 5, 0 | −3.38 | 75 |
| Middle temporal gyrus (BA 21) | L | −60, 0, −5 | −3.30 | 93 |
| Superior temporal gyrus (BA 38) | L | −55, 5, −10 | −3.30 | 69 |
| Inferior temporal gyrus (BA 20) | L | −55, −5, −35 | −3.30 | 54 |
| Middle temporal gyrus (BA 39) | L | −45, −80, 20 | −3.15 | 33 |
| **Parietal lobe** | | | | |
| Precuneus (BA 7) | L,R | −5, −35, 45 | −3.57 | 144 |
| Paracentral lobule (BA 5) | L,R | 0, −35, 50 | −3.54 | 52 |
| Precuneus (BA 31) | L,R | −10, −50, 35 | −3.53 | 35 |
| Superior parietal lobule (BA 7) | R | 25, −65, 45 | −3.35 | 33 |
| Postcentral gyrus (BA 3) | L | −20, −30, 50 | −3.24 | 48 |
| Superior parietal lobule (BA 7) | L | −25, −60, 45 | −3.20 | 34 |
| **Occipital lobe** | | | | |
| Middle occipital gyrus (BA 19) | L | −45, −85, 10 | −3.08 | 36 |
| **Limbic lobe** | | | | |
| Cingulate gyrus (BA 31) | L,R | 0, −35, 40 | −3.59 | 81 |
| Cingulate gyrus (BA 24) | L,R | 0, −25, 40 | −3.52 | 66 |
| Cingulate gyrus (BA 23) | L,R | 0, −25, 35 | −3.47 | 23 |
| Parahippocampal gyrus (BA 34) | L | −30, 5, −20 | −3.00 | 9 |
| Parahippocampal gyrus (BA 28) | L | −15, −5, −15 | −2.90 | 7 |
| **Sub-lobar** | | | | |
| Insula (BA 13) | L | −45, 5, 5 | −3.39 | 97 |
| Insula (BA 13) | R | 45, −5, 15 | −3.11 | 29 |

Corr. $p<0.05$ for $|t|>2.78$; BA – Brodmann areas; L – left; R – right; $x, y, z$ – MNI coordinates.

**Table 3.** Correlations between the changes in normalized upper alpha current source density for the Happy Memories with rtfMRI-EEG-nf conditions relative to the Rest conditions (H vs R) and anhedonia severity (SHAPS) ratings for the experimental group (EG).

| Region | Laterality | x, y, z (mm) | Corr. coeff. r | Voxels |
|---|---|---|---|---|
| **Frontal lobe** | | | | |
| Middle frontal gyrus (BA 9) | L | −30, 20, 35 | −0.744 | 34 |
| Inferior frontal gyrus (BA 9) | L | −35, 5, 30 | −0.732 | 19 |
| Middle frontal gyrus (BA 8) | L | −30, 15, 45 | −0.731 | 28 |
| Superior frontal gyrus (BA 8) | L | −20, 15, 50 | −0.712 | 30 |
| Precentral gyrus (BA 6) | L | −35, 0, 30 | −0.711 | 18 |
| Middle frontal gyrus (BA 6) | L | −30, 10, 50 | −0.710 | 25 |
| Middle frontal gyrus (BA 9) | R | 45, 30, 40 | −0.703 | 36 |
| Medial frontal gyrus (BA 6) | L,R | −5, 15, 50 | −0.703 | 30 |
| Middle frontal gyrus (BA 8) | R | 40, 30, 45 | −0.702 | 30 |
| Superior frontal gyrus (BA 6) | L | −20, 10, 55 | −0.699 | 37 |
| Superior frontal gyrus (BA 8) | R | 40, 20, 55 | −0.684 | 39 |
| Superior frontal gyrus (BA 6) | R | 5, 10, 55 | −0.680 | 36 |
| Inferior frontal gyrus (BA 47) | L | −50, 45, −10 | −0.678 | 24 |
| **Parietal lobe** | | | | |
| Inferior parietal lobule (BA 40) | R | 45, −35, 40 | −0.677 | 14 |
| **Limbic lobe** | | | | |
| Cingulate gyrus (BA 32) | L,R | −15, 15, 35 | −0.752 | 50 |
| Cingulate gyrus (BA 24) | L,R | −10, 15, 30 | −0.748 | 40 |
| **Sub-lobar** | | | | |
| Insula (BA 13) | L | −35, 5, 20 | −0.726 | 24 |

Corr. $p<0.05$ for $|r|>0.66$; BA – Brodmann areas; L – left; R – right; x, y, z – MNI coordinates.

**Table 4.** Changes in normalized high-beta current source density between the Happy Memories with rtfMRI-EEG-nf and Rest conditions (H vs R) for the experimental group (EG).

| Region | Laterality | x, y, z (mm) | t-score | Voxels |
|---|---|---|---|---|
| **Frontal lobe** | | | | |
| Medial frontal gyrus (BA 6) | L | −5, −15, 55 | −3.23 | 32 |
| Medial frontal gyrus (BA 6) | R | 5, −10, 55 | −3.19 | 24 |
| Superior frontal gyrus (BA 6) | L | −5, −5, 70 | −3.14 | 7 |
| Superior frontal gyrus (BA 6) | R | 5, 0, 70 | −3.13 | 8 |
| **Parietal lobe** | | | | |
| Paracentral lobule (BA 31) | L,R | −5, −15, 50 | −3.23 | 10 |
| **Limbic lobe** | | | | |
| Cingulate gyrus (BA 24) | L,R | −5, −10, 50 | −3.23 | 36 |

Corr. $p<0.05$ for $|t|>3.03$; BA – Brodmann areas; L – left; R – right; x, y, z – MNI coordinates.